\documentclass[reprint,showpacs,preprintnumbers,amsmath,amssymb,aip]{revtex4-2}

\usepackage{chemformula} 
\usepackage[T1]{fontenc} 
\usepackage{subcaption}
\usepackage{epstopdf}
\usepackage{ulem}

\begin{document}
\title{Ion Pair Dissociation Dynamics in an Aqueous Premelting Layer
}
\author{Samuel P. Niblett} 
\affiliation{Materials Science Division, Lawrence Berkeley National Laboratory, Berkeley, CA 94609 \looseness=-1}
\affiliation{Department of Chemistry, University of California, Berkeley CA 94609 \looseness=-1}
\email{sniblett@lbl.gov}

\author{David T. Limmer} 
\affiliation{Department of Chemistry, University of California, Berkeley CA 94609 \looseness=-1}
\affiliation{Kavli Energy NanoScience Institute, Berkeley, CA 94609 \looseness=-1}
\affiliation{Materials Science Division, Lawrence Berkeley National Laboratory, Berkeley, CA 94609 \looseness=-1}
\affiliation{Chemical Science Division, Lawrence Berkeley National Laboratory, Berkeley, CA 94609\looseness=-1}
\email{dlimmer@berkeley.edu}

\date{\today}
\begin{abstract}
  Using molecular dynamics simulations and methods of importance sampling, we study the thermodynamics and dynamics of sodium chloride in the aqueous premelting layer formed spontaneously at the interface between ice and its vapor. We uncover a hierarchy of timescales that characterize the relaxation dynamics of this system, spanning the picoseconds of ionic motion to the 10s-100s of nanoseconds associated with fluctuations of the liquid-crystal interface in their presence.
  We find that ions distort both local interfaces, incurring restoring forces that result in the ions preferentially residing in the middle of the layer.
  While ion pair dissociation is thermodynamically favorable,
  these structural and dynamic effects cause its rate to vary by over an order of magnitude through the layer, with a maximum rate significantly depressed from the corresponding bulk value. The solvation environment of ions in the premelting layer is distinct from that in a bulk liquid, being dominated by slow reorganization of water molecules and a water structure intermediate between ice and its melt.
\end{abstract}
\maketitle

\section*{Introduction}

The mechanisms and rates of reactions at extended interfaces can be dramatically different from those of the homogeneous materials that make them up.\cite{benjamin1996chemical,RuizLopezFMA20} This is especially true of reactions in water, as their mechanisms are often sensitive to the hydrogen bonding network, which is heavily disrupted by an interface. Considerable advances have been made in understanding reactions at water-air interfaces,\cite{Benjamin14,VenkateshwaranVG14,RuizLopezFMA20,GalibL20} but ice-air surfaces remain poorly studied despite their importance in the chemistry of the polar atmosphere.\cite{dash2006physics} Here, we use molecular dynamics simulations to study the paradigmatic model system of ion pair dissociation  within the premelting layer of ice.  We find that emergent structural and dynamic properties of the interface depress the overall dissociation rate relative to a bulk liquid, with a strong dependence on where in the layer the dissociation event occurs.

Reactions at ice interfaces contribute significantly to the chemical composition of the atmosphere.\cite{domine2002air}
 This influence arises from various settings and chemical cycles. Some reactions release atmospherically-active species, such as accelerated conversion of HOBr and HOCl into reactive Br$_2$/Cl$_2$ at the surface of ice particles in stratospheric clouds\cite{Abbatt94,BiancoH06} and on sea ice.\cite{Domine17} These halogen molecules subsequently photolyse into halogen radicals that contribute to ozone depletion. Other surface reactions remove trace gases, for example the oxidation of SO$_2$ to H$_2$SO$_4$ that contributes to ice and snow acidification.\cite{Abbatt03,ConklinSLV93} Finally, ice particles in urban settings catalyze photolysis of organic pollutant molecules,  often resulting in more toxic bi-products.\cite{KahanD07}

Understanding how ice-air interfaces affect chemical reactivity is therefore an important task, which remains challenging due to the complex structure of these interfaces. At typical polar temperatures, the surface of an ice crystal is covered by a thin film of disordered water molecules called the premelting layer or quasi-liquid layer.\cite{GoleckiJ78,ElbaumLD93} This layer forms to mitigate the large surface tension of an exposed ice crystal, since the liquid-like surface structure has fewer dangling hydrogen bonds and a smaller surface dipole.\cite{SlaterM19}
There are several outstanding questions surrounding premelting layers, particularly regarding their lateral extent and dynamics of formation.\cite{Limmer16,SlaterM19,qiu2018so, llombart2020surface} Here we consider how this complex, fluctuating environment affects reactivity.

\begin{figure*}[t]
    \includegraphics[width=15cm]{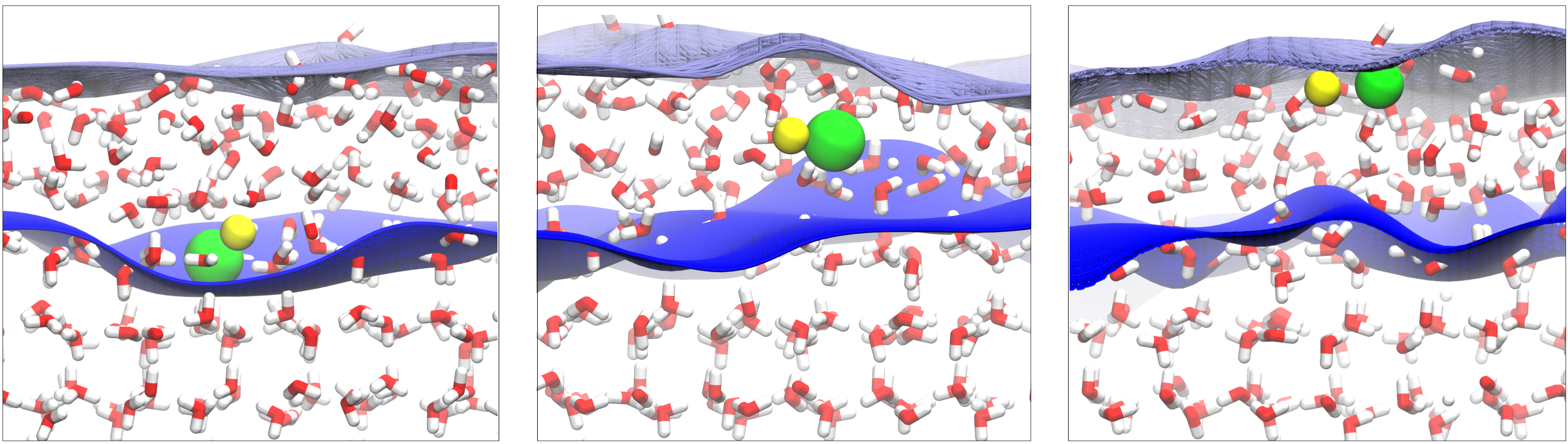}
  \caption{Typical configurations of a contact ion pair from molecular simulations. The sodium ion is shown in yellow, the chloride in green. Dark and light blue surfaces represent the liquid-crystal and liquid-vapor interfaces, respectively. From left to right, the three panels show the ionic centers of mass at $\bar{z}=0,\ell/2$ and $\ell$.}
  \label{fig:Model_geom}
\end{figure*}

Specifically, we probe the dynamics of a prototypical reaction, the dissociation of sodium chloride, in the aqueous premelting layer using classical simulation and theory.
Separation of singly-charged atomic ions has long been a preferred case study to understand solution-phase reaction dynamics computationally,\cite{JorgensenBHR87,GeisslerDC99,VenkateshwaranVG14} partly because the forces between the ions and solvent molecules are simple and easy to model, and partly because the dynamics of the reaction are intrinsically connected to those of the solvent.
It is now well-established that the dissociation reaction mechanism has a significant contribution from solvent degrees of freedom,\cite{GeisslerDC99} although how best to describe this contribution in a simple reaction coordinate remains a subject of debate.\cite{BallardD12,MullenSP14,Yonetani17,SalanneTVR17,RoyBMS17} NaCl dissociation is a particularly suitable test case for our purposes because the optimal reaction coordinate depends on water molecules up to 8\AA~ away from the ions themselves.\cite{BallardD12} This wide interaction sphere will be frustrated by the confines of the premelting layer, clearly demonstrating the dynamic effects of this environment.

The paper is organized as follows. We first introduce the model and simulation details used, giving particular attention to methods for identifying the boundaries of the premelting layer and discussion of the time scales exhibited by the system. We then discuss the equilibrium and energetic properties of the ion interface system, focusing on the response of the interface to the solutes and the free energy landscape experienced by the ions. Finally, we evaluate the rate of ion dissociation, revealing how the rate behavior depends on the local solvation environment, and interpret these results using reaction rate theory.

\section*{Simulating ice-vapor interfaces}
We studied ion pair dissociation in the premelting layer using molecular dynamics simulations. Water was represented by the TIP4P-2005 model,\cite{AbascalV05} sodium and chloride ions were modeled as charged Lennard-Jones atoms with interaction parameters taken from the OPLS-aa force field.\cite{JorgensenMT96} These forcefields have been shown to give reasonable descriptions of water solvation structure and thermodynamics close to the melting transition.\cite{FennellBVD09, BallardD12} 

Most of our calculations were performed using an Ih ice crystal containing 8x6x8 primitive unit cells with the crystallographic $c$ axis parallel to the $z$ direction. An additional vacuum region was added in contact with the basal (0001) plane of the crystal, increasing the length of the $z$ dimension to 170\AA. This slab geometry contains two ice-vapor interfaces, on the high- and low-$z$ faces of the crystal.  A single NaCl pair was equilibrated at the large-$z$ interface. This geometry is illustrated in Fig.~\ref{fig:Model_geom}. Periodic boundary conditions were applied in all directions, with short-range interactions truncated at 8.5\AA~ and electrostatic interactions computed by the particle mesh Ewald method.\cite{HockneyE88}

To obtain a nanometer-thick premelting layer, we performed our calculations at 2\,K below the ambient pressure melting temperature of TIP4P-2005. This implies a simulation temperature of 250K.\cite{AbascalV05} 
Prior to inserting the ion pair, the surface was equilibrated using a 50\,ns MD simulation where the temperature increased from 200K to 250K at a rate of 1K/ns, and a further 100\,ns simulation with a constant temperature of 250K. 
 Except where otherwise noted, a Langevin thermostat was used to control the temperature, with a damping time of 10\,ps where we wished to study dynamics, or 1ps if only static properties were being considered. Simulations were performed using the LAMMPS package.\cite{Plimpton95} 

To describe the premelting layer we require a means of identifying the interface between the liquid and crystal (lc), and that between the liquid and vapor (lv). On mesoscopic length scales, phases are defined through structural order parameters that manifest the globally broken symmetry associated with the phase transition. On molecular scales, an interface between two phases can be associated with the rapid spatial variation of a similar suitable local order parameter. Using the method of Willard and Chandler, we define interfaces using isosurfaces of a continuous coarse-grained order parameter field.\cite{WillardC10,LimmerC14}

The coarse-grained field is obtained by defining an appropriate local order parameter $\phi_\alpha(\mathbf{r})$ at field point $\mathbf{r}$ to distinguish the $\alpha =\{\rm{lv},\rm{lc}\}$ interface. Due to the discreteness of atoms and molecules, any function defined directly from atomic positions would be rapidly varying. Therefore we convolute $\phi_\alpha(\mathbf{r})$ with a Gaussian smoothing function,
\begin{equation}
  \overline{\phi}_\alpha(\mathbf{r};\xi) = \int {\rm d}\mathbf{r}' \, \phi_\alpha(\mathbf{r}') \frac{e^{-(\mathbf{r}-\mathbf{r}')^2/2\xi^2}}{(2\pi\xi)^{3/2}}  \, ,
\end{equation}
and arrive at a smoothly varying order parameter field whose variation can encode the location of long lived interfaces. The smoothing function is parameterized by a coarse-graining lengthscale $\xi$, chosen to be comparable to a molecular diameter. 

The construction of our model system ensures that both interfaces are roughly planar. Crystalline periodicity means that most properties of the system should be translationally invariant in the plane parallel to the interfaces, so for convenience we will write simulation coordinates as $\mathbf{r} = \{\mathbf{x},z\}$.
The interface $\alpha$ is defined locally at $\mathbf{x}$ as 
\begin{equation}
h_\alpha(\mathbf{x},t) =\max_z \, \delta \left [\phi_\alpha^{\,c} -\overline{\phi}_\alpha(\mathbf{x},z;\xi) \right ] \, ,
\end{equation}
where $\phi_\alpha^{\,c}= (\langle\overline{\phi}_\alpha\rangle_i+\langle\overline{\phi}_\alpha\rangle_j)/2$, 
$\langle \ldots\rangle_i$ denotes an average over phase $i$, and $\delta$ is Dirac's delta function.
The maximization over $z$ acts to select the upper interface, where the ions are located.
In practice, we evaluate the order parameter field on a regular cartesian grid and locate the isosurface by linear interpolation between gridpoints.

Further coarse-graining can provide information on longer length scales. We define a semi-local interface height
\begin{equation}
  H_\alpha(\mathbf{x},t) = \frac{1}{ \pi \Delta^2} \int  \mathrm{d}\mathbf{x}' \, \Theta(\Delta-|\mathbf{x}'-\mathbf{x}|) h_\alpha(\mathbf{x},t) \, ,
\end{equation}
where $\Theta$ is the Heaviside step function.  This height, $H_\alpha(\mathbf{x},t)$, is the instantaneous mean height of the $\alpha$ interface in a circle of radius $\Delta$ centered at $\mathbf{x}$. Both $h_\alpha(\mathbf{x},t)$ and  $H_\alpha(\mathbf{x},t)$ fluctuate in time as molecules change their configurations. 

For the liquid-vapor interface, the appropriate order parameter,
$
\phi_{\rm lv}(\mathbf{r}) =\sum_{i=1}^{N_\mathrm{w}} \delta(\mathbf{r}-\mathbf{r}_i) \, ,
$
 is just the local number density computable as a sum over all of the water molecules. Here $\mathbf{r}_i$ is the position vector of water molecule $i$'s center of mass. 
We adopted the parameters $\xi =$ 2.4\AA~ and $\phi_\mathrm{lv}^{\,c} =$ 0.016\AA$^{-3}$, which were previously shown to give an accurate and convenient definition of the lv-interface.\cite{WillardC10}
The liquid-crystal  interface requires a more complex order parameter. The 6th-degree Steinhardt-Nelson-Ronchetti order parameter is convenient for this use.\cite{SteinhardNR83, moore2010freezing,LimmerC14} This function is defined as
$
  \phi_{\rm lc}(\mathbf{r}) = \sum_{i=1}^{N_\mathrm{w}} q(\mathbf{r}_i)\delta(\mathbf{r}-\mathbf{r}_i) \, ,
$
where $q(\mathbf{r}_i)$ is a projection of the local density onto the 6th-degree spherical harmonic.
\begin{eqnarray}
 q(\mathbf{r}_i) &=&  \left( \sum_{m=-6}^6 \left| \sum_{j \in{\rm nn}(i)} \sum_{k\in{\rm nn}(j)} Y_{6m}(\phi_{jk},\theta_{jk})\right|^2\right)^{1/2} \, ,
\end{eqnarray}
 where $Y_{lm}(\phi_{ij},\theta_{ij})$ is the $l$-$m$ spherical harmonic function of the angular coordinates for vector $(\mathbf{r}_i-\mathbf{r}_j)$, and nn$(i)$ indicates the 4 nearest neighbors of atom $i$. 
 Previous work has shown\cite{LimmerC14} that appropriate parameters for the lc-interface are $\xi =$ 2.5\AA~ and $\phi_\mathrm{lc}^{\,c} =$0.0167 \AA$^{-3}$.

Figure~\ref{fig:Model_geom} depicts typical configurations of an ion pair in the premelting layer. The two interfaces separate the system into three clear phases as expected. Both interfaces are perpendicular to $z$ on average, but undergo substantial fluctuations out of that plane with mean squared deviations $\langle \delta h_\mathrm{lc}^2 \rangle \approx 2$\AA$^2$ and  $\langle \delta h_\mathrm{lv}^2\rangle \approx 1$\AA$^2$, where $\delta h = h - \langle h \rangle$. The premelting layer is defined as the volume enclosed by the lc and lv interfaces on an exposed surface of the ice block. The layer thickness is given by $\ell = \langle h_{\rm lv}(\mathbf{x};t)-h_{\rm lc}(\mathbf{x};t) \rangle$. In the following, we choose the origin $z=0$ such that $\langle h_\mathrm{lc}\rangle = 0$ at 250K in the absence of ions. The $z$ coordinates will be expressed as fractions of $\ell = \langle h_\mathrm{lv}\rangle \approx 7$\AA~ under the same conditions.

\section*{\label{sec:Timescales}Relaxation within the premelting layer}
\begin{figure}[t]
    \includegraphics[width=8.5cm]{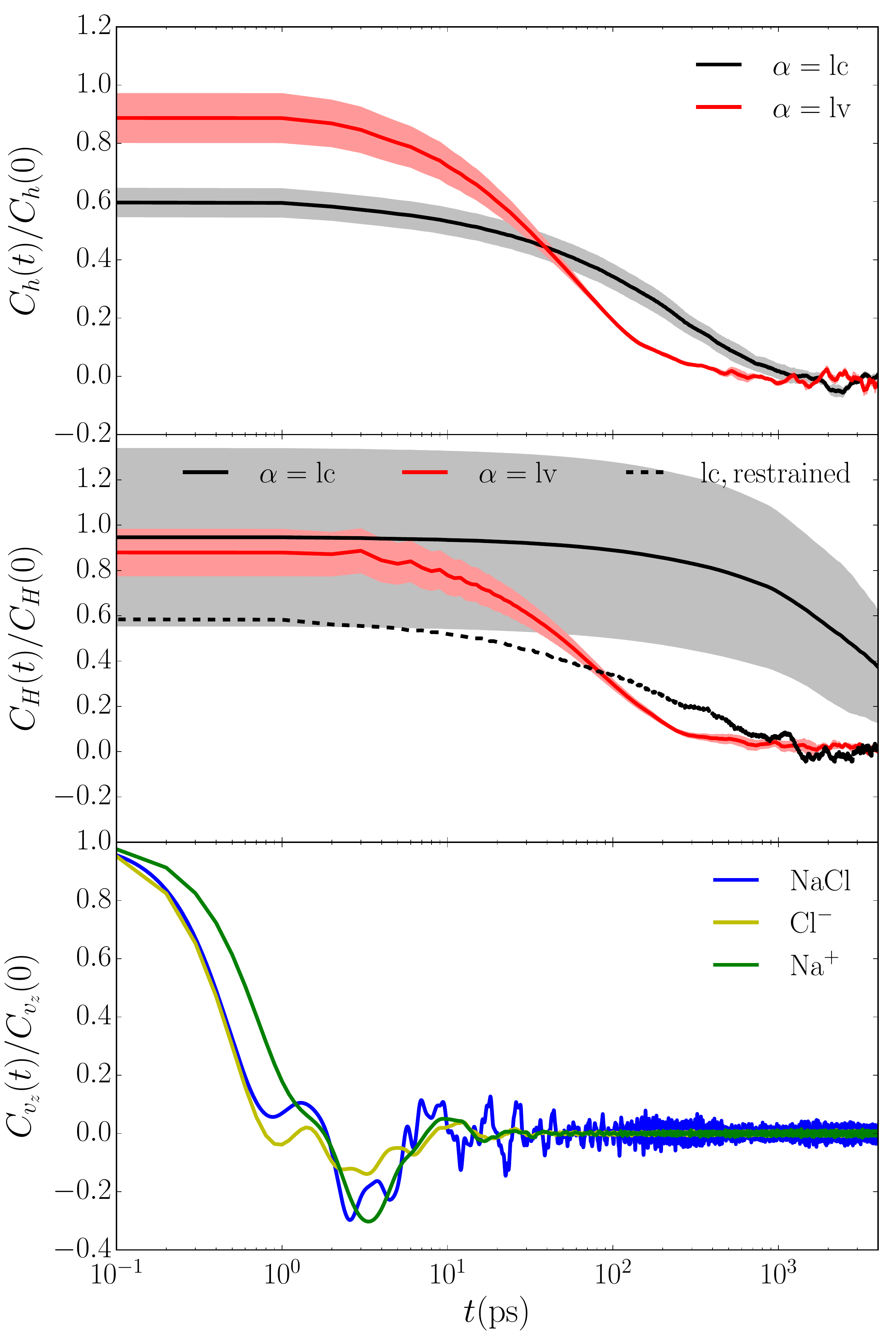}
    \caption{Relaxation timescales for the premelting layer. The top panel shows the autocorrelation function for the height of an individual site on the interface, $h_\alpha(\mathbf{x},t)$. The middle panel shows the autocorrelation of a spatially-averaged interface height $H_\alpha(\mathbf{x},t)$, with $\Delta=$1nm. In both panels, the shaded regions indicate the statistical error of the correlation function. The dashed line in the middle panel represents a simulation subject to a harmonic restraint on the ice molecules, as explained in the text. The last panel shows the autocorrelation function for the $z$ component velocity of different ionic species.}
  \label{fig:Timescales}
\end{figure}

The typical dynamics of ions within the premelting layer are marked by a hierarchy of distinct relaxation timescales. 
Figure~\ref{fig:Timescales} summarizes several measures of local relaxation within the premelting layer. Each panel shows time correlation functions for the deviation of a local quantity from its mean. The local height autocorrelation function 
\begin{equation}
C_{h}(t) = \langle \delta h_\alpha(\mathbf{x},0)\delta h_\alpha(\mathbf{x},t)\rangle \, ,
\end{equation} 
decays to 0.1 of its initial value within 0.1 ns for the lv-interface, and within 0.5 ns for the lc-interface. The faster decorrelation of the lv-interface is consistent with expectations from capillary wave theory in that relaxation times scale inversely proportional to the surface tension.\cite{ThakrePdOB08}

Figure~\ref{fig:Timescales} also shows the autocorrelation function of the local mean height 
\begin{equation}
C_{H}(t) = \langle \delta H_\alpha(\mathbf{x},0)\delta H_\alpha(\mathbf{x},t)\rangle \, ,
\end{equation}
for $\Delta=1$nm, representing a large enough radius to encompass the three solvation shells of the NaCl ion pair which are believed to influence dissociation dynamics.\cite{BallardD12} The lv-interface autocorrelation decays within approximately 0.1\,ns, comparable to the time required for a single site to decorrelate.
The lc-interface autocorrelation on the other hand, decorrelates much more slowly, at least 10 times slower than the corresponding single-site function. The average interface height retains some memory of its position for tens to hundreds of nanoseconds.
This slow decay is due to rare, discontinuous jumps in the mean interface height, where a layer of water molecules melts or freezes across large regions of the ice interface.\cite{sanchez2017experimental,llombart2020rounded} The long waiting time between jumps makes averaging over these fluctuations challenging, resulting in the large error estimate for $C_H(t).$ 

To mitigate the difficulty of studying the structure and dynamics of ions in the presence of slow diffusion of the lc-interface, in most of our analysis we restrained the water molecules located below $z=-0.4\ell$, attaching them to their equilibrium positions using a harmonic potential with spring constant $k=5$\,kCal\,\AA$^{-2}$. This spring constant permits vibrations of magnitude approximately 0.5\AA~but no diffusion of the ice molecules, so that melting is suppressed. The dashed line in the middle panel of fig.~\ref{fig:Timescales} demonstrates that height deviations of the lc-interface decay quickly in the absence of the melting event, recovering the behavior of $C_h(t)$. 
The $z=-0.4\ell$ limit ensures that the restraint does not affect molecules that are likely to come in contact with the ion pair, so they can still freeze and melt locally over the course of thermal fluctuations without nucleating global diffusion of the interface.

Finally, in Fig.~\ref{fig:Timescales} we consider the vertical motion of the ions. The $z$-directional velocity autocorrelation function, 
\begin{equation}
C_{v_z }(t) =\langle v_z(0)v_z(t)\rangle \, ,
\end{equation}
is largely decayed within 10ps for each ion individually, and for the center of mass of a bound ion pair. 
We computed the $z-$directional diffusivities of the ions from $C_{v_z }(t)$ using the Green-Kubo formula,\cite{Green54,Kubo57} obtaining values of $D_z= 7.4 \times10^{-8}$\,cm$^2$s$^{-1}$ for a free cation, 5.6 $\times10^{-8}$\,cm$^2$s$^{-1}$ for a free anion and $8.9 \times10^{-7}$\,cm$^2$s$^{-1}$ for the ion pair.
These diffusion constants are extremely slow, two orders of magnitude smaller than the ambient-temperature bulk diffusion constants.\cite{BenavidesPCEAV17}
indicating that the ionic species require approximately 10\,ns to diffuse over a distance equal to their molecular diameter. Such slow diffusivities imply a very sluggish dynamic environment, and illustrate the challenges of studying rare events in the premelting layer.

\begin{figure}
    \includegraphics[width=8.5cm]{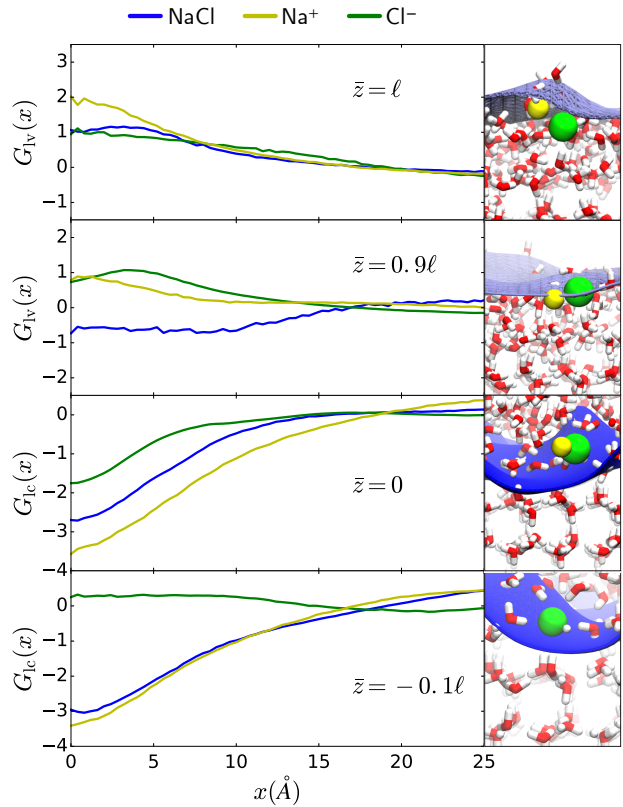}
    \caption{The average distortion of the interfaces, $G_\alpha(x)$, when one or two ions are nearby. Each panel shows results for a single sodium ion, single chloride ion and a sodium chloride pair constrained to their equilibrium separation. Different panels show different values of the ionic center-of-mass $\bar{z}$ coordinate. Alongside each panel is a corresponding typical configuration of the ions and nearest interface.}
  \label{fig:heights}
\end{figure}

\section*{Ion-interface spatial correlations}
While the dynamics of the ions and interfaces are not strongly correlated, we expect that the ions and interfaces will influence each other statically.
To quantify these spatial correlations, we examined the local shape of each interface near an ion. We calculated the mean deviation of the interface height from its average interface height,
\begin{equation}
G_\alpha(x)  = \langle h_{\alpha}(x) \rangle_{\mathbf{r}_i} - \langle h_{\alpha}(x) \rangle,
\end{equation}
conditioned on a solute present at $\mathbf{r}_i=\{0,\bar{z}_i\}$.
We consider the dependence of this correlation function on the magnitude $x=|\mathbf{x}|$. 
Capillary wave theory\cite{nelson2004statistical,LimmerC14,BenetLSM16} predicts long-ranged spatial fluctuations in the position of an interface, with no characteristic length scale. To check the impact of finite size effects from these fluctuations, we performed these calculations on a system size twice as large in each of the $\mathbf{x}$ directions. 

\begin{figure*}[t]
    \includegraphics[width=15.5cm]{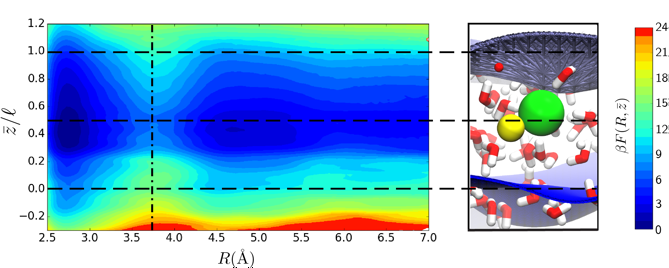}
    \caption{Free energy as a function of ionic separation $R$ and center-of-mass $z$ coordinate, $\bar{z}$. Three characteristic $\bar{z}$ values are indicated with dashed lines. The dash-dotted line approximately indicates the dissociation energy barrier, $R^\ddag(\bar{z})$.}
  \label{fig:pmf}
\end{figure*}
When $\bar{z}\approx \ell/2$ and the ions lie near the center of the premelting layer, the average deviation of the interface from its equilibrium position is negligible. Figure~\ref{fig:heights} shows how each interface responds when the ions approach.
In all panels $G_\alpha(x) \rightarrow 0$ at large $x$, showing that the influence of the ions has a finite range.  The correlation lengths associated with the height deviations are up to 1nm for the lv-interface and up to 2nm for the lc-interface, equivalent to several molecular diameters.

As the ions approach $\bar{z}=\ell$, the lc-interface retains its equilibrium statistics while the lv-interface locally rises by up to 2\AA. This increase results from water molecules being drawn upwards to cover the ions and screen their charge from the vacuum. Its magnitude generally increases with increasing $\bar{z}$.
A notable exception to this rule occurs at $\bar{z}=0.9\ell$, where the ion pair actually depresses $h_{lv}$ relative to its instantaneous position. At this particular $\bar{z}$, the ions are frequently sit parallel to the mean lv-interface and are left uncovered by the water. The ions themselves do not contribute to the interface calculation, so the uncovered ions lead to a depression in the interface.

Conversely, when the ions approach $\bar{z}=0$ the lv-interface is unaffected while the lc-interface is locally depressed. This depression indicates local disordering of the ice structure leading to improved ion solvation. The magnitude of the deviation varies with number of ions and $\bar{z}$, but usually falls between 1 and 4\AA. This magnitude is larger than the equivalent deviations in $h_{lv}$, indicating that the lc interface is softer and easier to deform consistent with its lower surface tension.

An ion pair or free sodium ion at $\bar{z}\approx-0.1\ell$ creates a significant depression in the lc-interface, but a free chloride ion does not. Instead, the anion incorporates into the ice surface structure, replacing one of the water molecules. Disruption of the periodic crystal structure is therefore minimized, and the height of the interface hardly changes. The cation is unable to incorporate in this way.
 Indeed in general, the influence of the cation on both interfaces is greater than that of the anion due to its greater charge density, which provides a strong bias to reorganize surrounding water molecules. With the exception of $\bar{z}=0.9\ell$, the effect of the ion pair is intermediate between the two free ions.

 None of the distortions observed in these calculations are significantly larger than the instantaneous capillary wave fluctuations for this system.
  The influence of ions appears to consist of pinning the equilibrium capillary waves in place. This pinning is associated with an entropic cost of moving the ions near to the interfaces, which influences the ion thermodynamics considerably\cite{OttenSGS12} as detailed below. 

\section*{Thermodynamics of ion pair dissociation}
To understand the thermodynamics of ion pairing in the premelting layer, we computed the free energy as a function of the ionic separation $R$ and center of mass vertical coordinate $\bar{z}$.
We employed umbrella sampling using a series of simulations with an additional harmonic potential, $V_b(R,\bar{z}) = k_r/2(R-R^\ast)^2 + k_z/2(\bar{z}-z^\ast)^2$. These simulations were combined using the weighted histogram analysis method\cite{GrossfieldWHAM} to compute an unbiased free energy, $F(R,\bar{z}) = -k_{\rm B}T {\rm ln}P(R,\bar{z})$, where $P(R,\bar{z})$ is the probability of observing the ions at separation $R$ and height $\bar{z}$ and $k_{\rm B}T$ is Boltzmann's constant times temperature. 
We used 478 distinct simulations, varying $R^\ast$ from 2.0\AA\, to 6.0\AA\, and $z^\ast$ from approximately -0.2$\ell$ to 1.0$\ell$. Half of the simulations used $k_r = k_z = 1$\,kCal/mol {\AA}$^2$ and the remainder employed stronger springs (up to 80kCal/mol \AA $^2$) to ensure adequate sampling. Simulation length was varied for the same reason, with most datasets representing 5\,ns of simulation time.

Figure~\ref{fig:pmf} shows the resulting free energy landscape, alongside a typical snapshot to illustrate its connection to the system geometry. For all $R$ values, the free energy is lowest near the middle of the premelting layer and increases monotonically towards each interface. 
This increase represents an interplay between reduced solvation, entropic penalties from pinning the interfaces, and the energy cost of disrupting the crystal structure.  Approaching the lv-interface, average coordination numbers for both bound and separated ion pairs decrease\cite{VenkateshwaranVG14} and the average electrostatic potential energy on the ions from the water increase in a manner anticipated from Fig.~\ref{fig:heights}a,b). Upon approaching the lc-interface, ion solvating water molecules reorganize, disrupting the ice structure and melting the crystal as anticipated from Fig.~\ref{fig:heights}c,d).

For all  $\bar{z}$ values, the free energy displays two clear basins of attraction, one for the contact ion pair (CIP) state in the range 2.5\,\AA<$R$<3.4\,\AA, and one for the solvent-separated pair (SSP) at around 4.5\,\AA<$R$<5.5\,\AA. These states are separated by a barrier that is approximately 6.5$k_{\rm B}T$ at $\bar{z}=0.5\ell$ and varies by up to 3$k_{\rm B}T$ over the premelting layer. The free energy also exhibits a broad, shallow minimum for $R$>6.5\,\AA, corresponding to completely dissociated ions. The energy barrier separating this minimum from the SSP state is very small. The free energy change to dissociate the ions in the premelting layer implies a six-fold decrease of the dissociation constant relative to bulk water around at the same temperature and pressure, based on an independent bulk free energy calculation.

\section*{Dynamics of ion pair dissociation}
We have shown that both static and dynamic fluctuations vary strongly through the premelting layer. Analogously, the rate of ion pair dissociation need not be constant and could similarly depend on where in the layer the rare event occurs.\cite{SchileL19} Since vertical motion of the ions is so slow, we will assume that dissociation events occur at fixed ionic $z$ coordinates. We have computed transition rates from the CIP to SSP as a continuous function of $\bar{z}$, using the Bennett-Chandler approach.\cite{Chandler78}

\begin{figure}[t]
    \includegraphics[width=8.5cm]{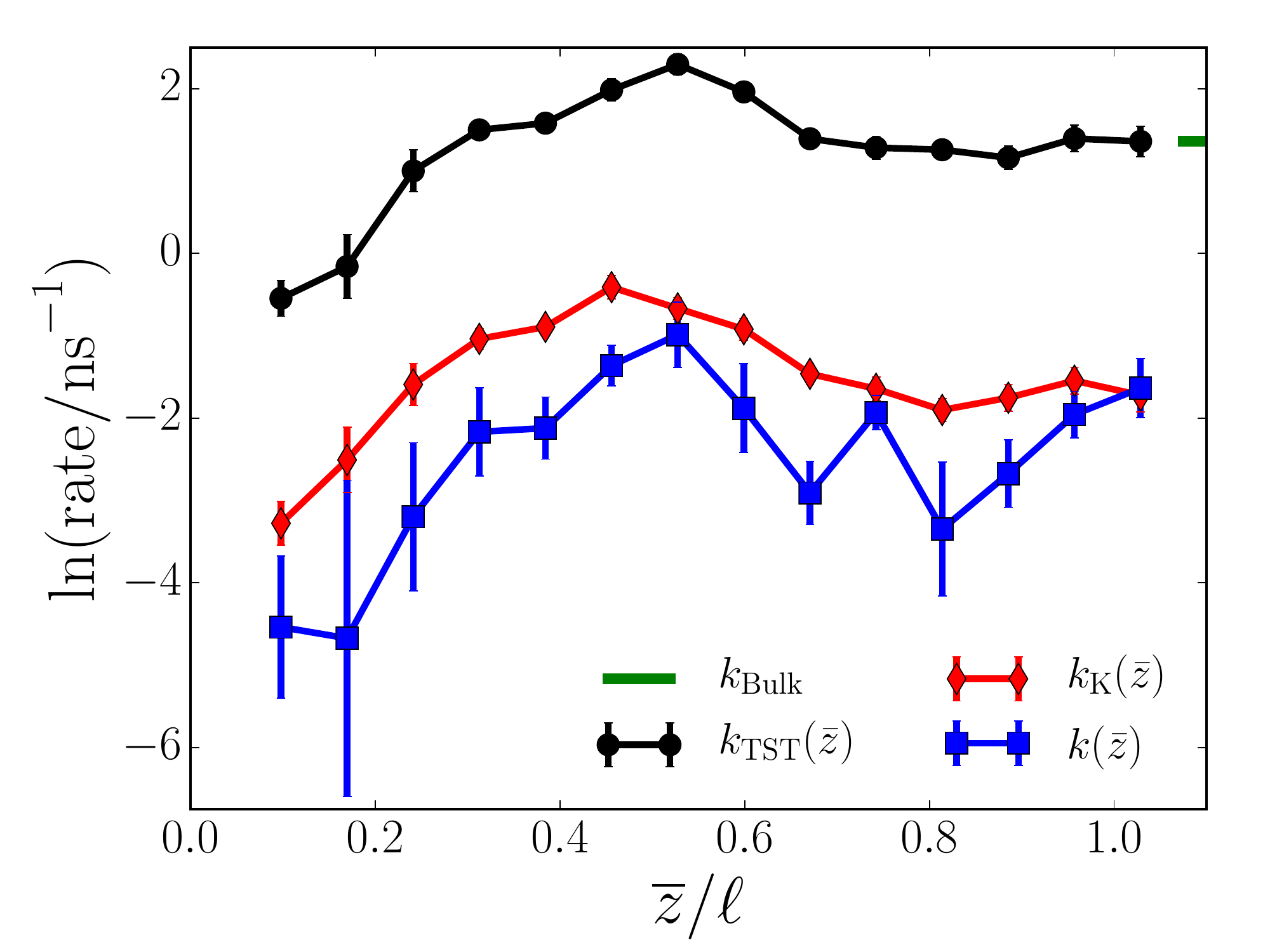}
    \caption{Variation of the ion dissociation rate with position in the premelting layer. Rate constants computed according to TST theory, Kramers theory and Bennett-Chandler theory are all shown. Lines shown are guides for the eye.}
  \label{fig:rates}
\end{figure}

We have defined the CIP reactant state and SSP product state using a dividing surface along the ionic separation distance, denoted $R^\ddag(\bar{z})$. For each fixed value of $\bar{z}$, $R^\ddag(\bar{z})$ is chosen as the location of the saddle point along $R$, which is close to $R=$3.7\,\AA~for all $\bar{z}$. The Bennett-Chandler rate is the product of a transition state theory estimate along this dividing surface, $k_{\rm TST}(\bar{z})$, and the transmission coefficient, $\kappa(\bar{z})$. The transition state theory estimate is the product of an attempt frequency and the probability of reaching the dividing surface from the reactant state
\begin{equation}
  \label{eq:TST}
    k_{\rm TST}(\bar{z})   = \frac{k_{\rm B}T}{2 \mu} \frac{e^{-\beta F(R^\ddag,\bar{z})}}{ \int_{-\infty}^{R^\ddag} ~{\rm d}R  \, e^{-\beta F(R,\bar{z})}} \, ,
\end{equation}
where $\mu$ is the reduced mass of the ion pair. 

The transmission coefficient is defined as the long time limit of the correlation function,
\begin{equation}
  \kappa(\bar{z}) =  \lim_{t \rightarrow \infty} \frac{\langle \dot{R}(0)\theta[R(t)-R^\ddag(\bar{z})]\rangle_\ddag}{\langle |\dot{R}(0)| \rangle_\ddag/2} \, ,
  \label{eq:kappa}
\end{equation}
which is the flux through the dividing surface conditioned on ending in the SSP state. The notation $\langle \cdots \rangle_\ddag$ indicates an average over an equilibrium ensemble of initial conditions for which $R = R^\ddag(\bar{z})$, obtained from constrained simulations. In practice this quantity is averaged independently for $\dot{R}(0)>0$ and $\dot{R}(0)<0$.  For each initial condition, an ensemble of initial velocities are generated from the Maxwell-Boltzmann distribution and corresponding trajectories are propagated for 10 ps with microcanonical trajectories, which is sufficient for the system to commit to either the reactant or product basin.
The ensemble averages in Eq.~(\ref{eq:kappa}) converge slowly, in our case requiring approximately 5000 trajectories. The transmission coefficients are very small, in the range 0.01-0.05 for the different $\bar{z}$ values. 

The dissociation rate,  $k(\bar{z}) = k_{\rm TST}(\bar{z}) \kappa(\bar{z})$, as a function of $\bar{z}$ is shown in Fig.~\ref{fig:rates}. The rate is slow near the lc-interface where the ions are essentially frozen in place. The rate increases towards the layer center, where improved solvation stabilizes the transition-state configurations, then drops off towards the lv-interface. The total variation over the premelting layer is over an order of magnitude. At its peak in the middle of the layer, the dissociation rate is an order of magnitude smaller than it would be in a bulk liquid at these same conditions, $k_\mathrm{bulk}$. This depression of the rate reflects how different the solvation environment is from a bulk solution, even in the middle of the liquid-like layer.

At $\bar{z}=0$ and below, the ions must either melt large regions of ice or incorporate into the crystal structure. Both processes are associated with slow and position-dependent dynamics. Moreover, describing ionic behavior in the crystal itself would probably require relaxing our harmonic restraint on the ice molecules and hence averaging our calculations over rare layer-melting events. These considerations meant that we were unable to obtain converged rate calculations for $\bar{z}\leq 0$. However, we note that the free energy cost of moving an ion pair this deep into the crystal is large.

We expect $R$ to be a poor descriptor of reaction dynamics in a premelting layer. Even in bulk water, previous investigations have found\cite{GeisslerDC99, MullenSP14,Yonetani17,SalanneTVR17} that solvent degrees of freedom must be included in the reaction coordinate to well describe the transition state ensemble. Nevertheless, the rate profile is largely determined by the $\bar{z}$ dependence of free energy to move along $R$, as encoded in the transition state theory estimate. This suggests that the recrossing dynamics do not greatly affect the competition between different environments for dissociation, despite depressing the rate by up to a factor of 100.

In order to understand the overall scale of the rate profile, we have compared $k(\bar{z})$ to that estimated from Kramers theory in the spatial diffusion limit.\cite{Kramers40,Peters17} In bulk water at ambient temperature, ion pair dissociation is known to have inertial components to its dynamics.\cite{MullenSP14} However, at low temperatures within the premelting layer small diffusivity of the ions in Fig.~\ref{fig:Timescales} suggest that that momentum correlations are damped out on the timescales of dissociation.  Assuming that the dynamics are overdamped, an approximate transmission coefficient can be evaluated in the harmonic barrier limit,
$\kappa_\mathrm{K}(\bar{z}) \approx \mu \omega_b D(R^\ddag,\bar{z})/k_{\rm{B}}T$, where $\omega_b$ is the imaginary frequency associated with the free energy barrier, and $D(R^\ddag,\bar{z})$ is the diffusion constant along $R$ evaluated at the top of the barrier at fixed $\bar{z}$. We estimate $D(R^\ddag,\bar{z})$ following a procedure due to Hummer, valid for simulations where the reaction coordinate $R$ is restrained to a particular window with a harmonic bias potential.\cite{Hummer05} Specifically, an autocorrelation function of $R(t)$ is computed under a harmonic bias to remain at the top of the barrier. Its value at $t=0$ and decay time can be used to compute $D(R^\ddag,\bar{z})$. Both values are obtained from 10ns constrained simulations using the biasing potentials employed in the umbrella sampling calculations. Consistent with our assumption that dissociation occurs at a fixed $\bar{z}$ value, we found that $D(R^\ddag,\bar{z}) \gg D_z$.
Therefore reactive events proceed much faster than vertical ionic motion, and the dissociation processes at values of $\bar{z}$ can be considered independent.\cite{berezhkovskii1989rate}

The Kramers rate, $k_\mathrm{K}(\bar{z})=k_\mathrm{TST}(\bar{z}) \kappa_\mathrm{K}(\bar{z})$, accounts for recrossing due to friction along the reaction coordinate. In Fig.~\ref{fig:rates}, we see that this estimate agrees fairly well with $k(\bar{z})$, particularly near the two interfaces.
This observation suggests that most of the observed recrossing is due to slow dynamics along the reaction coordinate, caused by the  sluggish environment of the premelting layer.
The biggest gap between the Kramers and Bennett-Chandler rate estimates comes in the range $0.1<\overline{z}<0.5$, perhaps indicating a more complex reaction mechanism associated with ions attaching to the ice crystal surface.

\section*{Conclusion}
The picture of the premelting layer that emerges from our simulations is that of a confined, sluggish environment, where water molecules adopt liquid-like configurations but with significant residual ice structure.
Both interfaces affect the dynamics of nearby reactions through energetic and dynamic factors, and in the case of ion dissociation both factors act to slow the reaction.
The ice interface appears to be much more significant for determining reaction dynamics, both in magnitude and range of interaction.

We  found that the ions significantly alter the properties of the nearby interfaces. The dynamics of both the lc- and lv-interfaces are dominated by capillary wave fluctuations decaying in approximately 1ns. When an ion pair resides near the interface these waves are pinned in place, creating a long-lived distortion of the interface shape. In the range of ion positions that are energetically feasible, the distortions do not exceed the equilibrium fluctuation size, but the entropic cost of pinning the wave contributes to the energy barrier which keeps the ions near the layer center.

The dissociation dynamics of water molecules in the premelting layer differ significantly from those in bulk solution, even at the layer center where the water structure appears most liquid-like.
Partially-dissociated configurations are harder to solvate in the premelting layer than contact ion pairs, leading to a significantly increased dissociation energy barrier.
Kinetics also play a more significant role in the premelting layer, mostly due to high friction from the slow water dynamics and residual ice-like structure. The implications of this slow, fluctuating environment on more complex reactions relevant in the polar regions are worthy of future study.

\section*{Acknowledgements}
This material is based on work supported by the U.S. Department of Energy, Office of Science, Office of Advanced Scientific Computing Research, Scientific Discovery through Advanced
Computing (SciDAC) program, under Award No. DE-AC02-05CH11231. This research used resources of the National Energy Research Scientific Computing Center (NERSC), a U.S. Department
of Energy Office of Science User Facility operated under Contract No. DE-AC02-05CH11231.


\section*{References}
\bibliography{ions}
\end{document}